\documentclass[reprint,amsmath,amssymb,aps,]{revtex4-1}
\usepackage{graphicx}
\usepackage{dcolumn}
\usepackage{amsmath}
\usepackage{amssymb}
\usepackage{bm}
\newcommand{\Tr}{\text{Tr}}
\begin{document}
\title{ Quantum Fisher information and Coherence of atom in dissipative cavity}
\author{Jianhe Yang}
\author{Danping Lin}
\author{Rongfang Liu}%
\author{Hong-Mei Zou}%
\email{zhmzc1997@hunnu.edu.cn}
\affiliation{Synergetic Innovation Center for Quantum Effects and Application, Key Laboratory of Low-dimensional Quantum Structures and Quantum Control of Ministry of Education, School of Physics and Electronics, Hunan Normal University, Changsha, 410081, P.R. China.}%

\begin{abstract}
	In this work, we investigate quantum Fisher information(QFI) and quantum coherence(QC) of an atom in dissipative cavity. In zero temperature reservior and with one excitation number, we obtain the analytical solutions of QFI and QC as well as their relationship for Ohmic and Lorentzian reservoir, respectively. The results show that both of the atom-cavity coupling and the cavity-reservoir coupling can effectively protect QFI and QC. Especially, QFI and QC will tend to their stable values when the atom-cavity coupling or the cavity-reservoir coupling is larger than a certain value. QC can augment QFI and can effectively improve the quantum metrology. In addition, we give a physical explanation of the dynamic behavior of QFI and QC by using the decoherence rate.
\begin{description}
\item[PACS numbers]
03.65.Yz, 03.67.-a, 42.50.Lc, 42.50.Pq
\item[Keywords]
quantum fisher information; coherence; dissipative cavity
\end{description}
\end{abstract}

\maketitle

\section{Introduction}
Quantum Fisher information(QFI) is a central concept in quantum metrology and quantum estimation theory, which effectively characterizes the statistical distinguishability about parameters encoded in quantum states and defines the precision of parameter estimation in quantum measurements through the Cram\'{e}r-Rao inequality\cite{Fisher,Helstrom,Wootters,Braunstein2,Dittmann,Barndorff,Paris,Giovannetti,Escher,Lu}. On the other hand, quantum coherence(QC) is one of important characteristics that quantum physics differs from classical physics. Quantum coherence constitutes powerful resources for quantum metrology\cite{Giovannetti2,Demkowicz} and entanglement creation\cite{Asboth,Streltsov}, and is widely used in quantum optics\cite{Glauber,Scully,Albrecht,Milburn}, quantum information\cite{Nielsen}.

As we known, any quantum system in the realistic world is open since it inevitably interacts with its environment. An open quantum system is governed by a nonunitary time development which describes features of irreversible dynamics such as the dissipation of energy and the decay of quantum coherence and correlation\cite{Alicki,Breuer}. QFI and QC of an open quantum system are inherently troubled by the interaction with surrounding environment. Recently, more and more attention has been paid to investigate QFI and QC of open quantum systems. For instance, the influence of an environment on QFI and quantum estimation has been extensively studied, including quantum Fisher information and quantum Fisher information flow of an open system, quantum metrology and parameter estimation precision in non-Markovian environment and quantum Fisher information of multiple phases in an open quantum system \cite{Lu2,Chin,Berrada2,YHHu,Zhong,H,Qiang1,Tan1,Ren1,Liao1}. Some schemes for measuring and protecting quantum coherence have also been proposed\cite{Baumgratz,Girolami,Thomas,Liu,Torun,Zhao,Liu2,Levi,Angel}. Moreover, one has studied the relationship between QFI and QC\cite{Feng,Wang}.

Much important progress has been obtained in theoretical and experimental researches on QFI and QC of open quantum systems, but these studies are mainly focused on some quantum systems interacting directly with their environments. In this paper, we investigate QFI and QC of an atom in a cavity interacting with a reservoir. We acquire analytical and numerical results of QFI and QC as well as their relationship when the cavity interacts with Ohmic and Lorentzian reservoir at zero temperature, respectively. The results show that, both of the atom-cavity coupling and the cavity-reservoir coupling can effectively protect QFI and QC.

The outline of the paper is the following. In Section 2, we provide a physical model. In Section 3, we obtain the expressions of QFI and QC. We discuss in detail the influence of the atom-cavity coupling and the cavity-reservoir coupling on QFI and QC in Section 4. Finally, we conclude with a brief summary of important results in Section 5.

\section{ Physical model}
We consider an atom in a single mode cavity coupling with a multi-mode reservoir. Assume the atom-cavity of interest is a part of a larger system whose dynamics is unitary, the Hamiltonian of the total system is given by
\begin{equation}\label{EB301}
\begin{split}
\hat{H}_{tot}=&\frac{1}{2}\omega_{0}\hat{\sigma}_{z}++\omega_{0}\hat{a}^{\dag}\hat{a}+\Omega(\hat{a}\hat{\sigma}_{+}+\hat{a}^{\dag}\hat{\sigma}_{-})\\
&+\sum_{k}\omega_{k}\hat{c}_{k}^{\dag}\hat{c}_{k}
+(\hat{a}^{\dag}+\hat{a})\sum_{k}g_{k}(\hat{c}_{k}^{\dag}+\hat{c}_{k}).
\end{split}
\end{equation}
In Eq.(\ref{EB301}), the first three correspond to the form of the Hamiltonian of the Jaynes-Cummings model\cite{Jaynes,Shore} and the last two describe the cavity leakage because the actual cavity is often not perfect and there is a leakage of the field from the cavity, which is usually modeled by coupling of the cavity mode to the bosonic modes of the reservoir\cite{Scala,Scala1}.
$\hat{\sigma}_{z}$ and $\hat{\sigma}_{\pm}$ are the Pauli matrices in the space of the atomic excited($|e\rangle$) and ground($|g\rangle$) states. $\hat{c}_{k}^{\dag}$($\hat{c}_{k}$) is the creation(annihilation) operator of the reservoir and $g_{k}$ is the coupling strength of the cavity-reservoir. $\hat{a}^{\dag}$($\hat{a}$) is the creation(annihilation) operator of the cavity and $\Omega$ is the coupling strength of the atom-cavity\cite{Jaynes}.

Under the Born-Markov and the rotating wave approximations, tracing out the reservoir degrees of freedom for the total density operator, assuming that the total system has only one initial excitation and the reservoir is at zero temperature in the following section. Let the dressed states are $|\alpha_{1\pm}\rangle=(|1g\rangle\pm|0e\rangle)/\sqrt{2}$ and $|\alpha_{0}\rangle=|0g\rangle$. The transition operators are $\hat{b}_{1}^{+}=|\alpha_{1,-}\rangle\langle\alpha_{0}|$, $\hat{b}_{1}^{-}=|\alpha_{0}\rangle\langle\alpha_{1,-}|$, $\hat{b}_{2}^{+}=|\alpha_{1,+}\rangle\langle\alpha_{0}|$ and  $\hat{b}_{2}^{-}=|\alpha_{0}\rangle\langle\alpha_{1,+}|$. The master equation for the density operator $\varrho(t)$ of the atom-cavity system\cite{Shore, Scala1,Zou3} is
\begin{eqnarray}\label{EB304}
\frac{d}{dt}\varrho(t)&=&-i[\hat{\mathcal{H}},\varrho(t)]
+\frac{1}{2}\gamma_{1}(t)(\hat{b}_{1}^{-}\varrho(t)\hat{b}_{1}^{+}-\frac{1}{2}\{\hat{b}_{1}^{+}\hat{b}_{1}^{-},\varrho(t)\})\nonumber\\
&+&\frac{1}{2}\gamma_{2}(t)(\hat{b}_{2}^{-}\varrho(t)\hat{b}_{2}^{+}-\frac{1}{2}\{\hat{b}_{2}^{+}\hat{b}_{2}^{-},\varrho(t)\}),
\end{eqnarray}
where $\hat{\mathcal{H}}=\frac{1}{2}\omega_{0}\hat{\sigma}_{z}+\omega_{0}\hat{a}^{\dag}\hat{a}+\Omega(\hat{a}\hat{\sigma}_{+}
+\hat{a}^{\dag}\hat{\sigma}_{-})$. $\gamma_{1}(t)$ and $\gamma_{2}(t)$ are the time-dependent decay rates for $|\alpha_{1,-}\rangle$ and $|\alpha_{1,+}\rangle$, respectively.

By means of the representation transformation and tracing out the cavity degrees of freedom, we can acquire the density matrix $\rho(t)$ of the atom in the standard basis$\{|e\rangle, |g\rangle\}$. Suppose the initial state is $|\Phi(0)\rangle=\cos\frac{\theta}{2}|e\rangle+e^{i\phi}\sin\frac{\theta}{2}|g\rangle$, $\rho(t)$ can be expressed as
\begin{eqnarray}\label{EB314}
\rho(t)=\left(
            \begin{array}{cc}
              |p(t)|^{2}\cos^{2}\frac{\theta}{2}& p(t)e^{-i\phi}\sin\frac{\theta}{2}\cos\frac{\theta}{2}\\
              p(t)^{\ast}e^{i\phi}\sin\frac{\theta}{2}\cos\frac{\theta}{2}&1-|p(t)|^{2}\cos^{2}\frac{\theta}{2}\\
              \end{array}
          \right),
\end{eqnarray}
where the probability amplitude $p(t)$ is written as
\begin{eqnarray}\label{EB306}
p(t)&=&\frac{1}{2}\sum_{j=1}^{2}e^{-i\omega_{j}t}e^{-\frac{1}{4}\beta_{j}},
\end{eqnarray}
in which $\omega_{1}=\omega_{0}-\Omega$ $(\omega_{2}=\omega_{0}+\Omega)$ is the transition frequency of the dressed-states $|\alpha_{1,-}\rangle\rightarrow|\alpha_{0}\rangle$ ($|\alpha_{1,+}\rangle\rightarrow|\alpha_{0}\rangle$), and
\begin{eqnarray}\label{EB307}
\beta_{j}&=&\int_{0}^{t}\gamma_{j}(t')dt',
\end{eqnarray}
here
\begin{eqnarray}\label{EB308}
\gamma_{j}(t)&=&2Re[{\int_{0}^{t}d\tau \int_{-\infty}^{+\infty}d\omega'e^{i(\omega_{j}-\omega')\tau}J(\omega')}],
\end{eqnarray}
when $J(\omega')$ is the spectral density of reservoir.

We rewrite Eq.~(\ref{EB314}) as a time-local master equation\cite{Breuer}, namely
\begin{eqnarray}\label{EB309}
\frac{d}{dt}\rho(t)&=&L\rho(t)\nonumber\\
&=&-\frac{i}{2}S(t)[\hat{\sigma}_{+}\hat{\sigma}_{-},\rho(t)]+\Gamma(t)\{\hat{\sigma}_{-}\rho(t)\hat{\sigma}_{+}\nonumber\\
&-&\frac{1}{2}\hat{\sigma}_{+}\hat{\sigma}_{-}\rho(t)-\frac{1}{2}\rho(t)\hat{\sigma}_{+}\hat{\sigma}_{-}\},
\end{eqnarray}
where
\begin{eqnarray}\label{EB310}
S(t)=-2Im[\frac{\dot{p}(t)}{p(t)}],
\end{eqnarray}
is the Lamb frequency shift to describe the contribution from the unitary evolution part under dynamical decoherence, and
\begin{eqnarray}\label{EB3101}
\Gamma(t)=-2Re[\frac{\dot{p}(t)}{p(t)}],
\end{eqnarray}
represents the decoherence rate of the atom to characterize the backflow of information from the environment to the atom.

\section{ Quantum Fisher information and Quantum coherence}
For quantum Fisher information, there are many different and useful versions, here we will adopt the one based on the symmetric logarithmic derivative\cite{Helstrom,Braunstein2,H,Qiang1}. $\phi$ denotes a single parameter to be estimated and $\mathcal{L}_{\phi}$ is symmetric logarithmic derivatives for the parameter $\phi$, QFI is defined as
\begin{equation}\label{EB21}
\mathcal{F}_{\phi}=\Tr(\rho \mathcal{L}_{\phi}^{2}),
\end{equation}
where $\mathcal{L}_{\phi}$ is determined by $\frac{\partial}{\partial\phi}\rho=\frac{1}{2}(\rho \mathcal{L}_{\phi}+\mathcal{L}_{\phi}\rho)$. Usually, it is difficult to give QFI in the form of the density operator for a general system. Fortunately, the QFI of the two-dimensional density matrix has obtained explicitly in Refs.\cite{Dittmann,Zhong} as
\begin{eqnarray}\label{EB25}
\mathcal{F}_{\phi}&=&\Tr[(\partial_{\phi}\rho)^{2}]+\frac{1}{\det(\rho)}\Tr[(\rho\partial_{\phi}\rho)^{2}].
\end{eqnarray}

Then adopting Eq.~(\ref{EB314}) and Eq.~(\ref{EB25}) by some straightforward calculation, the expressions of the QFI with respect to $\phi$ and $\theta$ are respectively acquired as
\begin{eqnarray}\label{EB315}
\mathcal{F}_{\phi}(t)&=&|p(t)|^{2}\sin^{2}\theta,\nonumber\\
\mathcal{F}_{\theta}(t)&=&|p(t)|^{2}.
\end{eqnarray}

From Eqs.~(\ref{EB315}), we know that $\mathcal{F}_{\phi}$ is dependent on $\theta$ and $p(t)$ while the characteristic of $\mathcal{F}_{\theta}$ is only determined by the function $p(t)$. There is $\mathcal{F}_{\phi}=\mathcal{F}_{\theta}$ when $\theta=\frac{\pi}{2}$.

Due to the fundamental importance of quantum coherence, a number of coherence measures have been proposed, such as the $l_{1}$ norm and relative entropy of coherence\cite{Baumgratz} and the skew information\cite{Girolami}. Here, we make use of the $l_{1}$ norm of coherence. From Eq.~(\ref{EB314}) we can get the expressions of the QC as
\begin{eqnarray}\label{EB316}
\mathcal{C}_{l}(t)=\sum_{i,j=1(i\neq j)}^{2}|\rho_{ij}(t)|=|p(t)\sin\theta|,
\end{eqnarray}
where $\rho_{ij}(t)(i\neq j)$ is the non-diagonal element of the density matrix $\rho(t)$. Eq.~(\ref{EB316}) shows that the $\mathcal{C}_{l}(t)$ is dependent on $\theta$ and $p(t)$.

Comparing Eqs.~(\ref{EB315}) and Eq.~(\ref{EB316}), we can obtain the analytical relationships between QFI and QC as
\begin{eqnarray}\label{EB317}
\mathcal{C}_{l}(t)&=&\sqrt{\mathcal{F}_{\phi}(t)},\nonumber\\
\mathcal{C}_{l}(t)&=&\sqrt{\mathcal{F}_{\theta}(t)\sin^{2}\theta}.
\end{eqnarray}

From Eqs.~(\ref{EB317}), we know that the quantum coherence can augment the QFI and can effectively improve the quantum metrology because the QFI will enlarge with the quantum coherence enlarging.

\section{Results and Discussions}
\subsection{ Ohmic reservoir with a Lorentz-Drude cutoff function}
If the reservoir is an Ohmic spectral density with a Lorentz-Drude cutoff function
\begin{equation}\label{EB321}
J(\omega')=\frac{2\omega'}{\pi}\frac{\omega_{c}^{2}}{\omega_{c}^{2}+\omega'^{2}},
\end{equation}
where $\omega'$ is the frequency of the reservoir, and $\omega_{c}$ is the cut-off frequency depending on the coupling strength. The case $\omega_{c}\ll\omega_{0}$ is essentially strong cavity-reservoir coupling. However, the case $\omega_{c}\gg\omega_{0}$ is weak cavity-reservoir coupling \cite{Sinayskiy,Eckel,CuiW}. Inserting Eq.~(\ref{EB321}) into Eq.~(\ref{EB308}), $\gamma_{j}(t)$ is
\begin{eqnarray}\label{EB322}
\gamma_{j}(t)&=&\frac{4\omega_{c}^2}{\omega_{j}^2+\omega_{c}^2}[\omega_{j}(1-e^{-w_{c}t}\cos(\omega_{j} t))\nonumber\\
&-&\omega_{c}e^{-w_{c}t}\sin(\omega_{j}t)],
\end{eqnarray}
Utilizing Eq.~(\ref{EB307}) and Eq.~(\ref{EB322}), the $\beta_{1}$ and $\beta_{2}$ have the following forms
\begin{eqnarray}\label{EB323}
     \beta_{1}&=&\frac{4\omega_{c}^2}{[(\omega_{0}-\Omega)^2+\omega_{c}^2]^2}\{[(\omega_{0}-\Omega)^2
     +\omega_{c}^2](\omega_{0}-\Omega) t\nonumber\\
     &+&2\omega_{c}(\omega_{0}-\Omega)[e^{-w_{c}t}\cos((\omega_{0}-\Omega)t)-1]\nonumber\\
     &-&[(\omega_{0}-\Omega)^2-\omega_{c}^2]e^{-w_{c}t}\sin((\omega_{0}-\Omega)t)\},\nonumber\\
     \beta_{2}&=&\frac{4\omega_{c}^2}{[(\omega_{0}+\Omega)^2+\omega_{c}^2]^2}\{[(\omega_{0}+\Omega)^2
     +\omega_{c}^2](\omega_{0}+\Omega) t\nonumber\\
     &+&2\omega_{c}(\omega_{0}+\Omega)[e^{-w_{c}t}\cos((\omega_{0}+\Omega)t)-1]\nonumber\\
     &-&[(\omega_{0}+\Omega)^2-\omega_{c}^2]e^{-w_{c}t}\sin((\omega_{0}+\Omega)t)\}.
\end{eqnarray}

Inserting Eq.~(\ref{EB323}) into Eq.~(\ref{EB306}) and then using Eq.~(\ref{EB315}) and Eq.~(\ref{EB316}), we can get $\mathcal{F}_{\phi}$ and $\mathcal{C}_{l}$.

Figure 1 describes the dynamics behavior of QFI and QC in the different $\Omega$ value. The dotted red line is for $\Omega=0.01\omega_{0}$ value, the dashed blue line is for $\Omega=0.5\omega_{0}$ and the solid green line is for $\Omega=\omega_{0}$. From figure 1(a), one can find that, in the weak cavity-reservoir coupling($\frac{\omega_{c}}{\omega_{0}}=3$), $\mathcal{F}_{\phi}$ will monotonously decrease to zero if $\Omega$ is very small but $\mathcal{F}_{\phi}$ reduces and then recovers to a stable value if $\Omega=\omega_{0}$. Figure 1(b) shows that, in the strong cavity-reservoir coupling($\frac{\omega_{c}}{\omega_{0}}=0.3$), $\mathcal{F}_{\phi}$ will slowly and monotonously decrease to zero when $\Omega=0.01\omega_{0}$. But an obvious oscillation occurs in the dynamics evolution of $\mathcal{F}_{\phi}$ as $\Omega$ increases and $\mathcal{F}_{\phi}$ tends a stable value when $\Omega=\omega_{0}$. Figure 1(c)-(d) draw the dynamical curves of $\mathcal{C}_{l}$ in the weak and strong cavity-reservoir coupling for different $\Omega$ value. Similar to the case of figure 1(a), $\mathcal{C}_{l}$ monotonously decays to zero when $\Omega$ is very small but $\mathcal{C}_{l}$ reduces and then recovers to a stable value when $\Omega=\omega_{0}$. Besides, the decay rate of $\mathcal{C}_{l}$ is smaller than that of $\mathcal{F}_{\phi}$. The curves of figure 1(d) is also similar to that of figure 1(b), the difference is also that the decay rate of $\mathcal{C}_{l}$ is smaller than that of $\mathcal{F}_{\phi}$.

\begin{center}
\includegraphics[width=4.2cm,height=3.5cm]{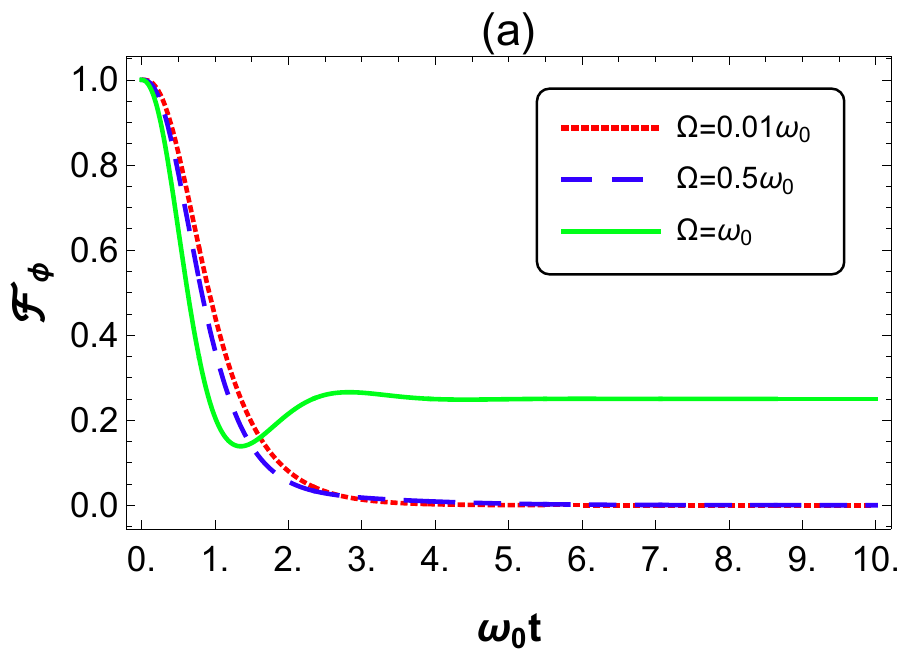}
\includegraphics[width=4.2cm,height=3.5cm]{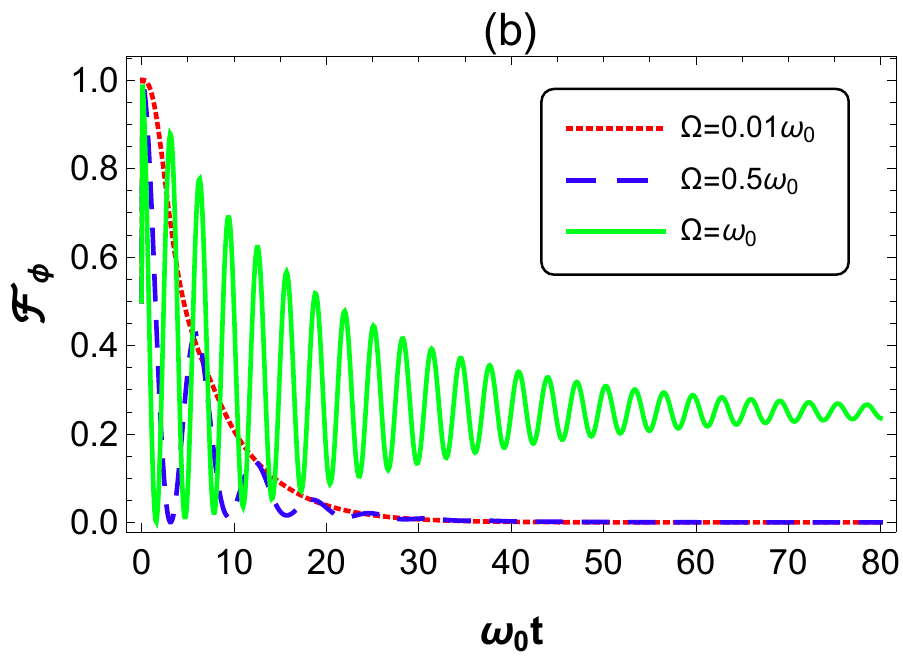}
\includegraphics[width=4.2cm,height=3.5cm]{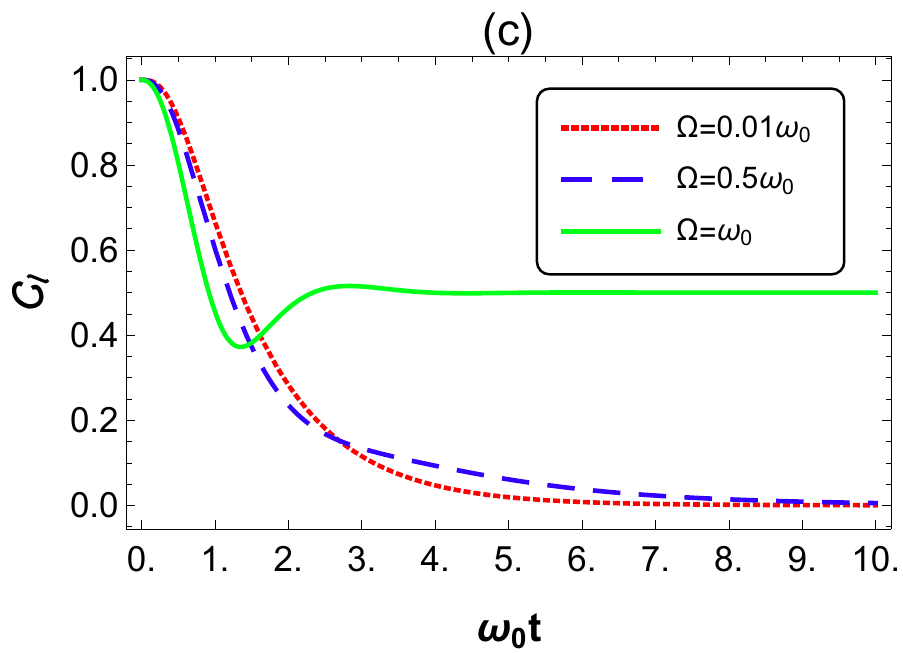}
\includegraphics[width=4.2cm,height=3.5cm]{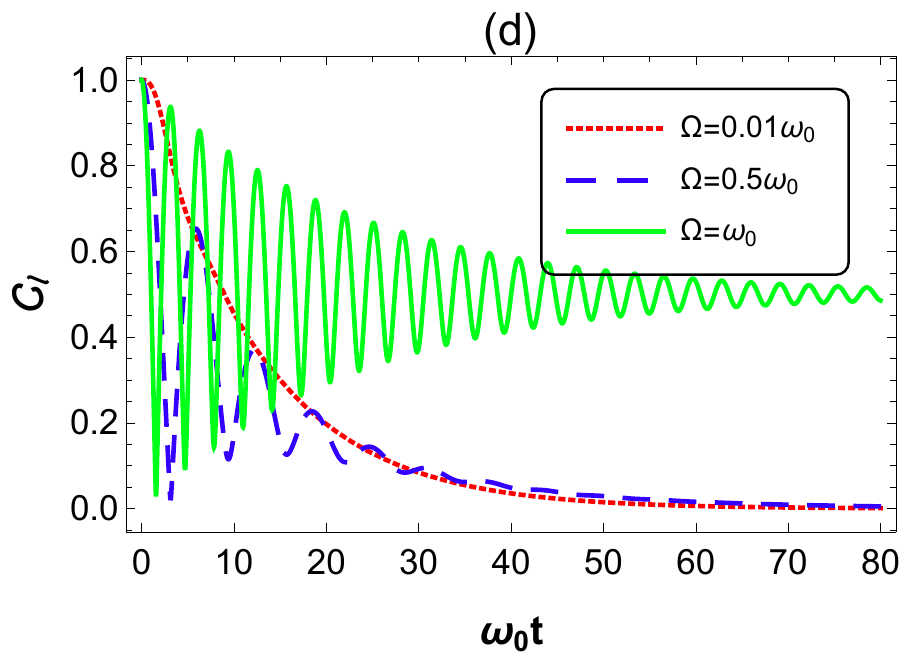}
\parbox{8cm}{\small{\bf Figure 1.}
(Color online)Quantum Fisher information and quantum coherence as a function of $\omega_{0}t$. $\Omega=0.01\omega_{0}$(red,dotted), $\Omega=0.5\omega_{0}$(blue, dashed) and $\Omega=\omega_{0}$(green, solid).  In the weak cavity-reservoir coupling($\frac{\omega_{c}}{\omega_{0}}=3$): (a)$\mathcal{F}_{\phi}$ as a function of $\omega_{0}t$ and (c) $\mathcal{C}_{l}$ as a function of $\omega_{0}t$. In the strong cavity-reservoir coupling($\frac{\omega_{c}}{\omega_{0}}=0.3$): (b)$\mathcal{F}_{\phi}$ as a function of $\omega_{0}t$ and (d) $\mathcal{C}_{l}$ as a function of $\omega_{0}t$. Other parameters are $\phi=0$ and $\theta=\frac{\pi}{2}$.}
\end{center}

\begin{center}
\includegraphics[width=4.2cm,height=3.8cm]{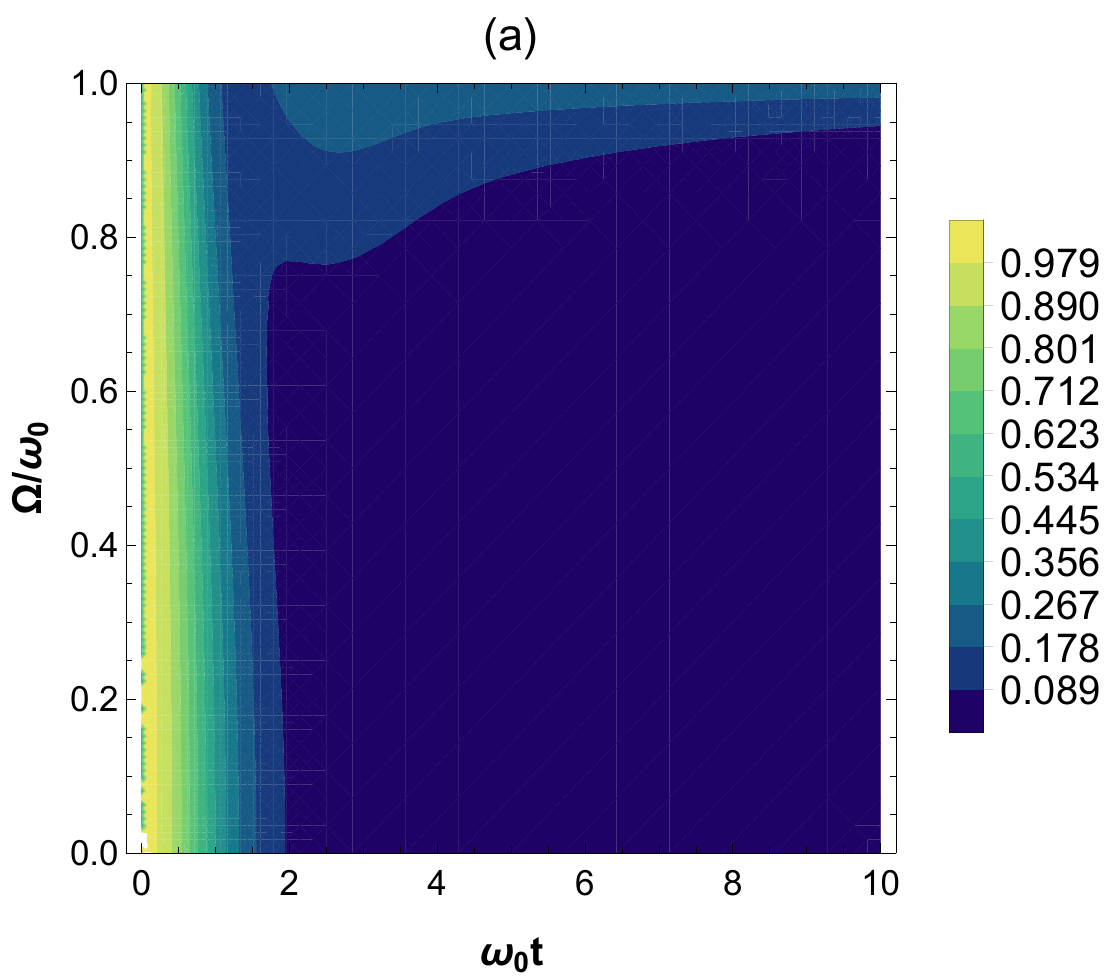}
\includegraphics[width=4.2cm,height=3.8cm]{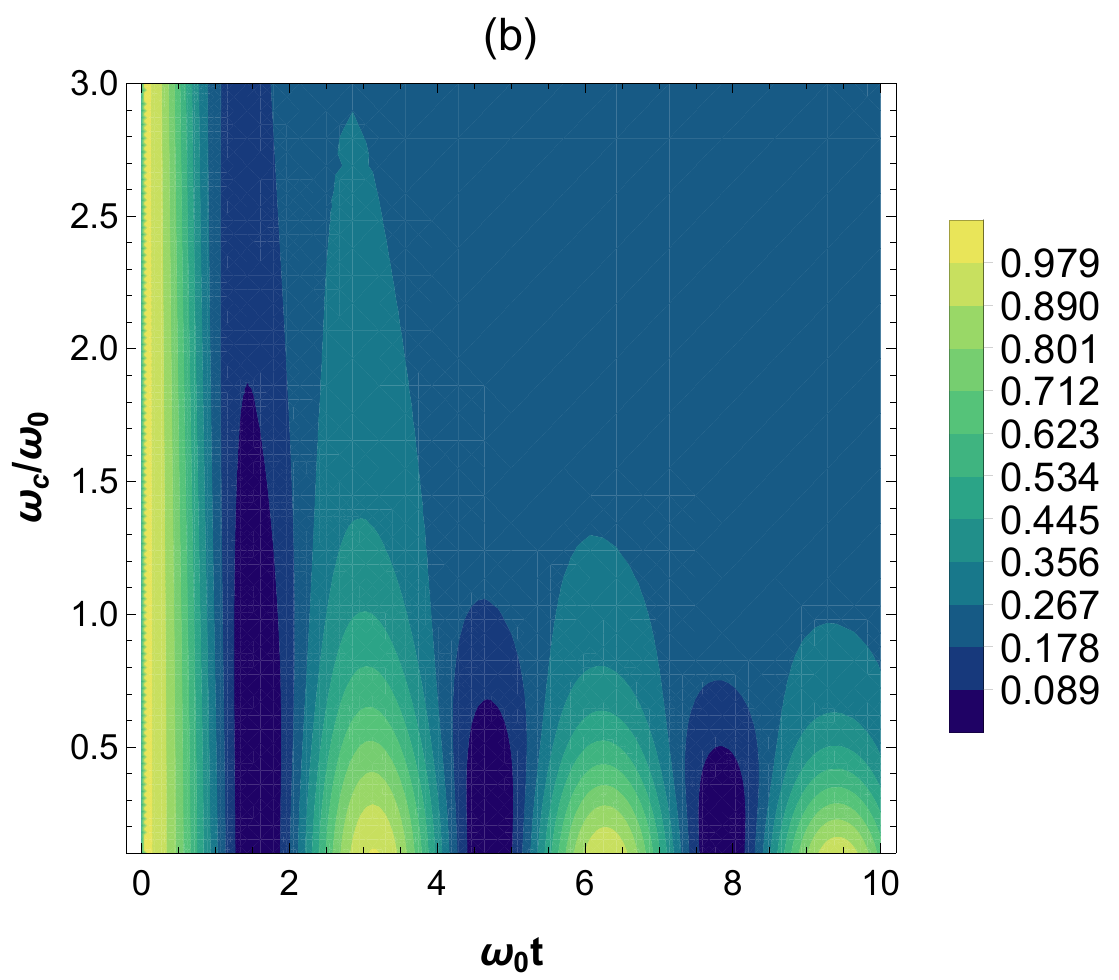}
\parbox{8cm}{\small{\bf Figure 2.}
(Color online)Contour map of quantum Fisher information. (a)$\mathcal{F}_{\phi}$ as a function of $\omega_{0}t$ and $\frac{\Omega}{\omega_{0}}$ when $\frac{\omega_{c}}{\omega_{0}}=3$. (b)$\mathcal{F}_{\phi}$ as a function of $\omega_{0}t$ and $\frac{\omega_{c}}{\omega_{0}}$ when $\Omega=\omega_{0}$.}
\end{center}

In figure 2(a), we plot $\mathcal{F}_{\phi}$ as a function of $\omega_{0}t$ and $\frac{\Omega}{\omega_{0}}$ in the weak cavity-reservoir coupling($\frac{\omega_{c}}{\omega_{0}}=3$). We can see that $\mathcal{F}_{\phi}$ decays monotonously when $\Omega$ is smaller however $\mathcal{F}_{\phi}$ oscillates damply to a stable value when $\Omega$ closes to $\omega_{0}$. Figure 2(b) exhibits $\mathcal{F}_{\phi}$ as a function of $\omega_{0}t$ and $\frac{\omega_{c}}{\omega_{0}}$ when $\Omega=\omega_{0}$. The result shows that $\mathcal{F}_{\phi}$ oscillates damply to a stable value in both of the weak and strong cavity-reservoir coupling, but the smaller the value of $\frac{\omega_{c}}{\omega_{0}}$ is, the more obvious the oscillation is. Furthermore, the evolution behavior of $\mathcal{C}_{l}(t)$ is similar to the $\mathcal{F}_\phi$, we omit it in order to reduce the space.

Thence, both of the atom-cavity coupling and the cavity-reservoir coupling can effectively protect the QFI and the QC. Namely, the enhancement of parameter estimation precision may be acquired by adjusting the atom-cavity coupling and the cavity-reservoir coupling. Besides, we see that $\mathcal{F}_{\phi}$ is equal to the square of $\mathcal{C}_{l}$ from their stable values, as the same as Eq.~(\ref{EB317}).

\begin{center}
\includegraphics[width=4.2cm,height=3.5cm]{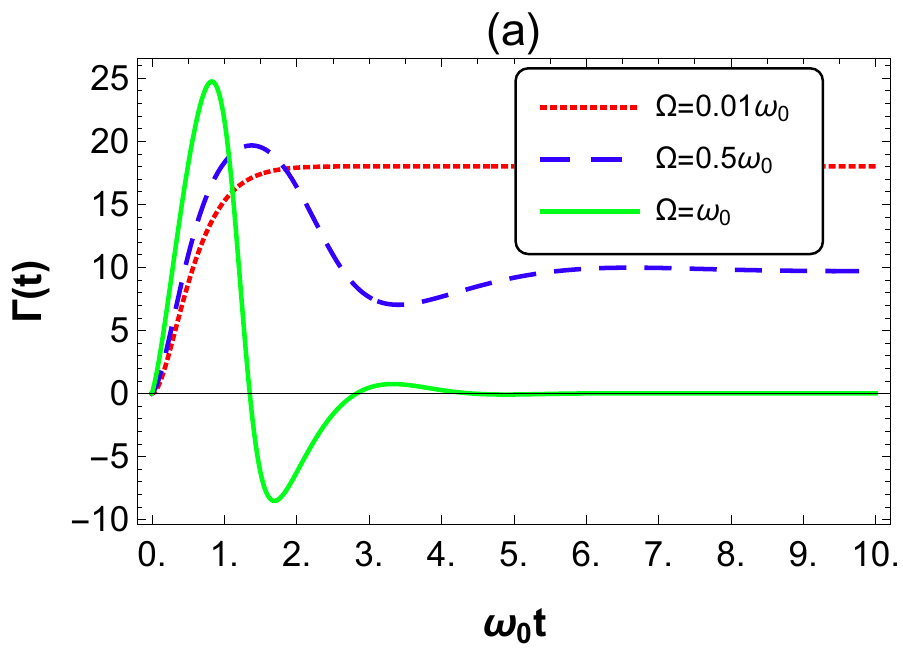}
\includegraphics[width=4.2cm,height=3.5cm]{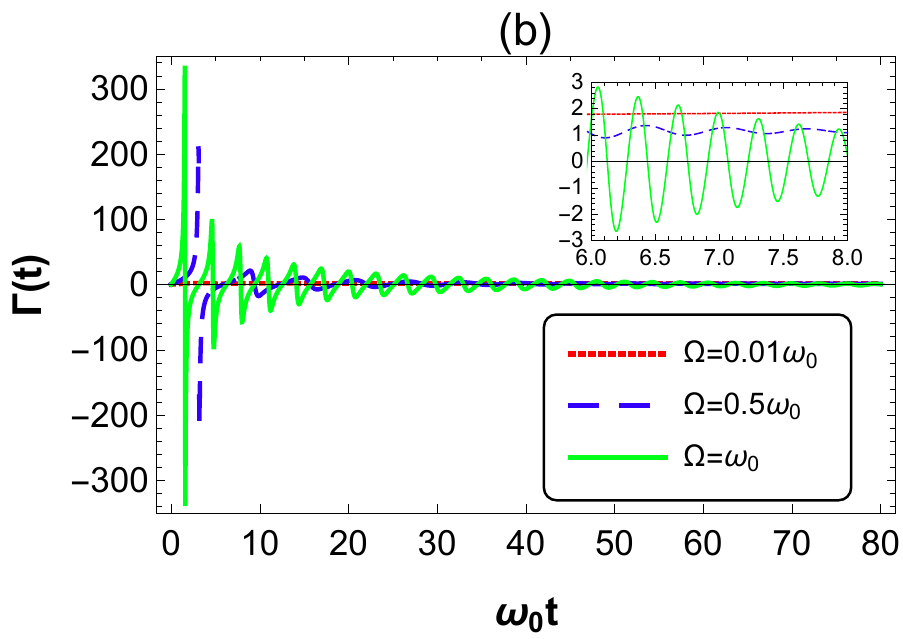}
\parbox{8cm}{\small{\bf Figure 3.}
(Color online)The decoherence rate $\Gamma(t)$ versus $\omega_{0}t$. $\Omega=0.01\omega_{0}$(red,dotted), $\Omega=0.5\omega_{0}$(blue, dashed) and $\Omega=\omega_{0}$(green, solid). (a)$\Gamma_{t}$ as a function of $\omega_{0}t$ in the weak cavity-reservoir coupling($\frac{\omega_{c}}{\omega_{0}}=3$); (b)$\Gamma_{t}$ as a function of $\omega_{0}t$ in the strong cavity-reservoir coupling($\frac{\omega_{c}}{\omega_{0}}=0.3$). Other parameters are $\phi=0$ and $\theta=\frac{\pi}{2}$.}
\end{center}

In order to explain the above results, we give the decoherence rate $\Gamma(t)$ as a function of $\omega_{0}t$ for different $\Omega$ value in figure 3. From Eq.~(\ref{EB301}), we know that the information and energy are exchanged between the atom with the cavity and between the cavity with the reservoir, thus the cavity and the reservoir are all regarded as the environment of the atom. The decoherence rate of the atom $\Gamma(t)$ depends on the coupling $\Omega$, the frequency $\omega_{0}$ of the cavity mode and the cut-off frequency $\omega_{c}$ of the reservoir mode, as shown in Eqs.~(\ref{EB306}), ~(\ref{EB3101}) and ~(\ref{EB323}). The case $\Gamma(t)>0$ indicates that the information flows irreversibly from the atom to the environment, but the case $\Gamma(t)<0$ shows that the information flows back from the environment to the atom. In the weak cavity-reservoir coupling($\frac{\omega_{c}}{\omega_{0}}=3$)(figure 3(a)), if $\Omega\leq0.5\omega_{0}$, the information flows irreversibly from the atom to the environment due to the dissipation of reservoir. $\Gamma(t)$ is always positive and $\mathcal{F}_{\phi}$ and $\mathcal{C}_{l}$ reduce quickly to zero with $\omega_{0}t$. If $\Omega=\omega_{0}$, the information will be fed back to the atom from the cavity. $\Gamma(t)$ changes from positive to negative values and then tends to zero so that $\mathcal{F}_{\phi}$ and $\mathcal{C}_{l}$ reduce and then recover to their stable values. In the strong cavity-reservoir coupling($\frac{\omega_{c}}{\omega_{0}}=0.3$)(figure 3(b)), if $\Omega\geq0.5\omega_{0}$, the information will flow back from the environment to the atom due to the memory and feedback effect of reservoir. $\Gamma(t)$ becomes larger and alternates between positive and negative values with $\Omega$ increasing so that $\mathcal{F}_{\phi}$ and $\mathcal{C}_{l}$ will oscillate significantly with $\omega_{0}t$. When $\Omega=\omega_{0}$, $\Gamma(t)$ will tend to zero thus $\mathcal{F}_{\phi}$ and $\mathcal{C}_{l}$ tend to their stable values rather than zero.

\subsection{ Lorentzian reservoir}
If the reservoir is a Lorentzian spectral density
\begin{equation}\label{EB311}
J(\omega')=\frac{1}{2\pi}\frac{R\lambda^{2}}{(\omega_{0}-\omega'-\Omega)^{2}+\lambda^{2}},
\end{equation}
where $\Omega$ is the detuning between the atom frequency $\omega_{0}$ and the center frequency $\omega'$ of the spectrum. The parameter $\lambda$ defines the spectral width of the coupling and the parameter $R$ is a dissipative rate. The case $\lambda>2R$ means strong cavity-reservoir coupling. The case $\lambda<2R$ is weak cavity-reservoir coupling \cite{Zou1,Bellomo1}. We can write $\gamma_{j}(t)$ from Eq.~(\ref{EB308}) and Eq.~(\ref{EB311}) as
\begin{eqnarray}\label{EB312}
\gamma_{j}(t)&=&\frac{R\lambda^{2}}{(\omega_{0}-\omega_{j}-\Omega)^{2}+\lambda^{2}}\nonumber\\
&&\{1+(\frac{\omega_{0}-\omega_{j}-\Omega}{\lambda}\sin((\omega_{0}-\omega_{j}-\Omega)t)\nonumber\\
&-&\cos((\omega_{0}-\omega_{j}-\Omega)t))e^{-\lambda t}\}.
\end{eqnarray}
Inserting Eq.~(\ref{EB312}) into Eq.~(\ref{EB307}), $\beta_{j}$ is stated as
\begin{eqnarray}\label{EB313}
\beta_{1}&=&R(t+\frac{e^{-\lambda t}-1}{\lambda}),\nonumber\\
\beta_{2}&=&\frac{R\lambda^{2}}{4\Omega^{2}+\lambda^{2}}[t-\frac{4\Omega e^{-\lambda t} sin(2\Omega t)}{4\Omega^{2}+\lambda^{2}}\nonumber\\
&+&\frac{(\lambda^{2}-4\Omega^{2})(e^{-\lambda t}cos(2\Omega t)-1)}{\lambda(4\Omega^{2}+\lambda^{2})}].
\end{eqnarray}

Inserting Eq.~(\ref{EB313}) into Eq.~(\ref{EB306}) and then using Eq.~(\ref{EB315}) and Eq.~(\ref{EB316}), we have $\mathcal{F}_{\phi}$ and $C_{l}$.

\begin{center}
	\includegraphics[width=4.2cm,height=3.5cm]{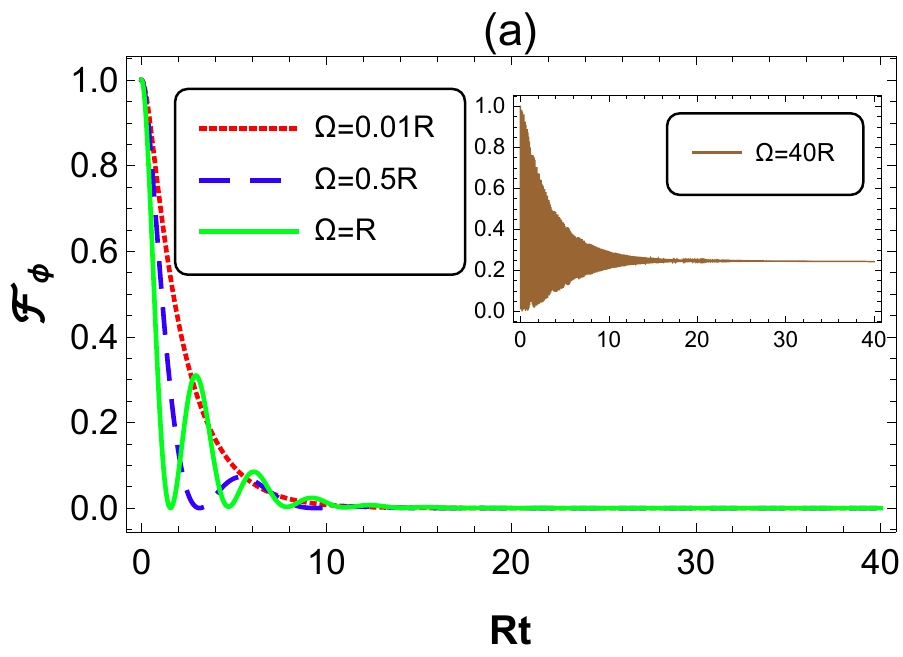}
	\includegraphics[width=4.2cm,height=3.5cm]{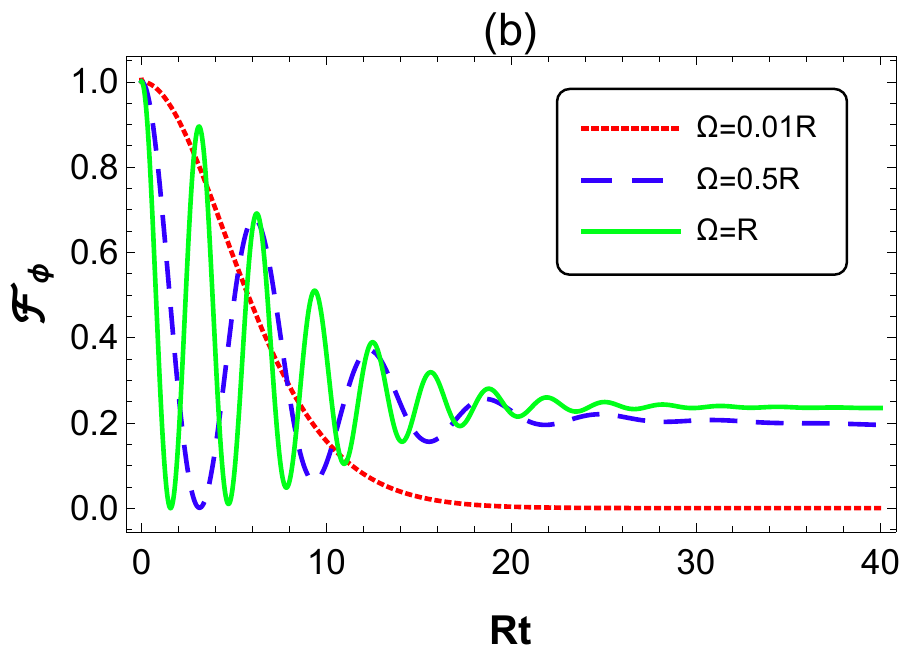}
	\includegraphics[width=4.2cm,height=3.5cm]{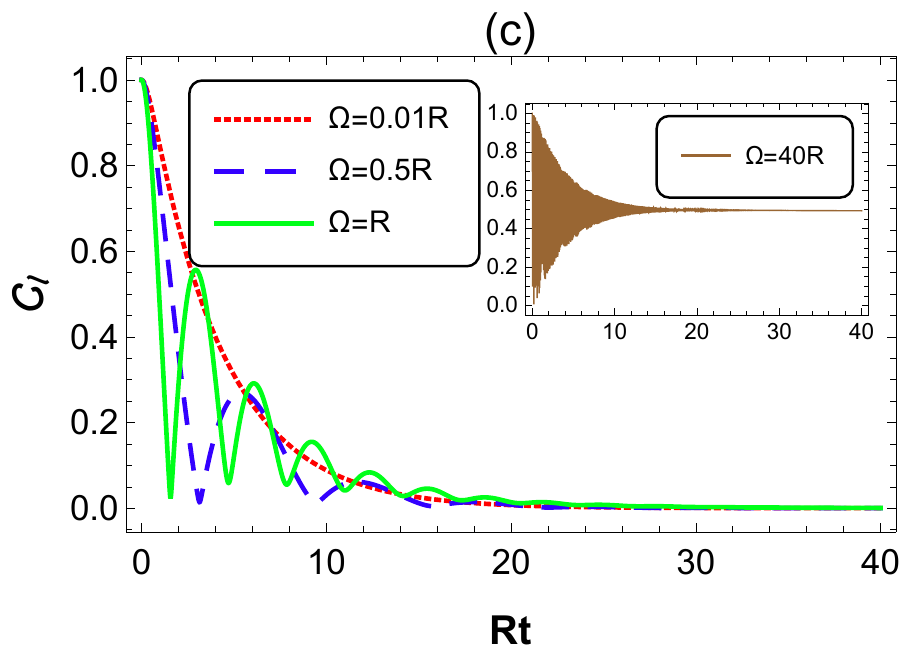}
	\includegraphics[width=4.2cm,height=3.5cm]{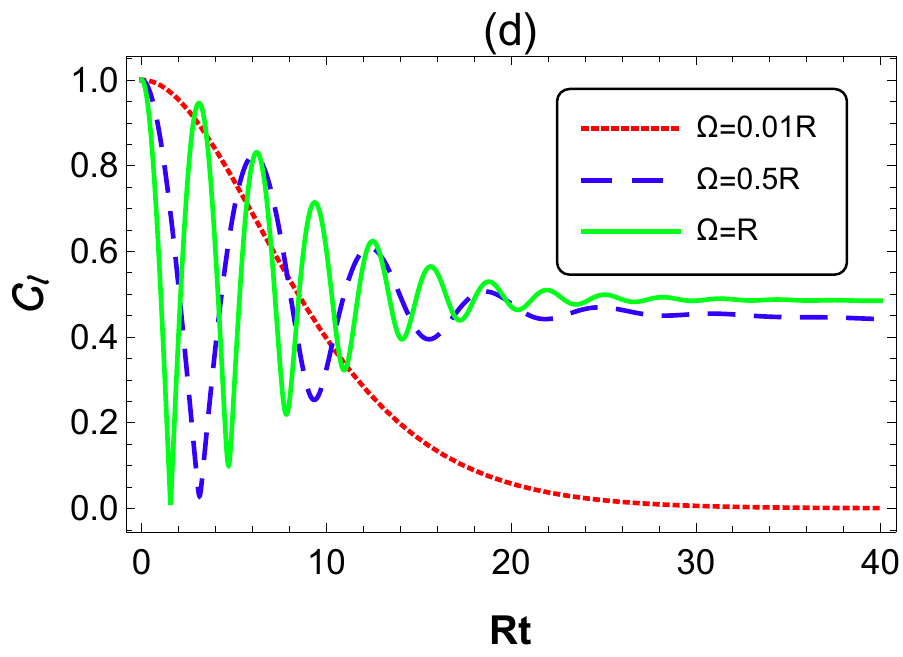}
	\parbox{8cm}{\small{\bf Figure 4.}
		(Color online)Quantum Fisher information and quantum coherence as a function of $Rt$. $\Omega=0.01R$(red,dotted), $\Omega=0.5R$(blue, dashed) and $\Omega=R$(green, solid).  In the weak cavity-reservoir coupling($\lambda=3R$): (a)$\mathcal{F}_{\phi}$ as a function of $Rt$ and (c) $\mathcal{C}_{l}$ as a function of $Rt$. In the strong cavity-reservoir coupling($\lambda=0.1R$): (b)$\mathcal{F}_{\phi}$ as a function of $Rt$ and (d) $\mathcal{C}_{l}$ as a function of $Rt$. Other parameters are $\phi=0$ and $\theta=\frac{\pi}{2}$.}
\end{center}

Figure 4 expresses the dynamical behaviour of QFI and QC in the different $\Omega$ value. The dotted red line is for $\Omega=0.01R$, the dashed blue line is for $\Omega=0.5R$, and the solid green line is for $\Omega=R$. In the weak cavity-reservoir coupling($\lambda=3R$)(figure 4(a)), one can see that $\mathcal{F}_{\phi}$ goes down exponentially to zero with $Rt$ when $\Omega=0.01R$. However, $\mathcal{F}_{\phi}$ will oscillate very obviously with $Rt$ and then decay to zero when $\Omega=0.5R$ and $\Omega=R$. We also find that the oscillation of $\mathcal{F}_{\phi}$ becomes faster and the decay of $\mathcal{F}_{\phi}$ becomes slower with $\Omega$ increasing. In particular, $\mathcal{F}_{\phi}$ will be close to a stable value when $\Omega=40R$(see the inset in figure 4(a)). Figure 4(b) shows the dynamical evolution of $\mathcal{F}_{\phi}$ in the strong cavity-reservoir coupling($\lambda=0.1R$) for different $\Omega$ values. Comparing figure 4(b) with figure 4(a), we see that $\mathcal{F}_{\phi}$ decays more slowly in the strong cavity-reservoir coupling than in the weak cavity-reservoir coupling and $\mathcal{F}_{\phi}$ can be close to a stable value when $\Omega=R$ in the strong cavity-reservoir coupling. Figure 4(c)-(d) display the dynamical behaviour of $\mathcal{C}_{l}$ in the weak and strong cavity-reservoir coupling for different $\Omega$ value. Similar to the case of figure 4(a), $\mathcal{C}_{l}$ monotonously decays to zero with $Rt$ when $\Omega=0.01R$, but $\mathcal{C}_{l}$ will oscillate damply to zero when $\Omega$ is small non-zero value and $\mathcal{C}_{l}$ is close to a stable value when $\Omega=40R$. But the decay rate of $\mathcal{C}_{l}$ is smaller than that of $\mathcal{F}_{\phi}$. The curves of figure 4(d) are also similar to that of figure 4(b), the difference is that the decay rate of $\mathcal{C}_{l}$ is smaller than that of $\mathcal{F}_{\phi}$.

\begin{center}
\includegraphics[width=4.2cm,height=3.8cm]{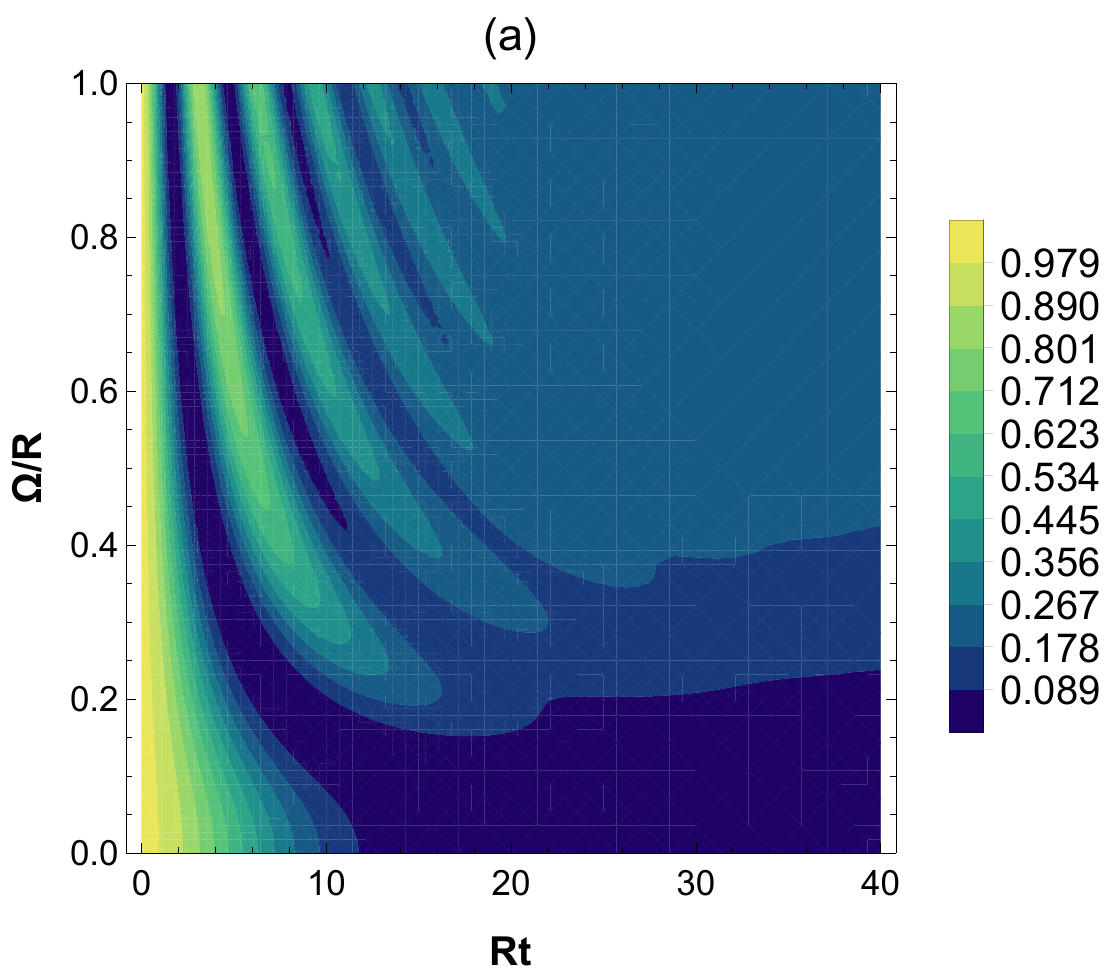}
\includegraphics[width=4.2cm,height=3.8cm]{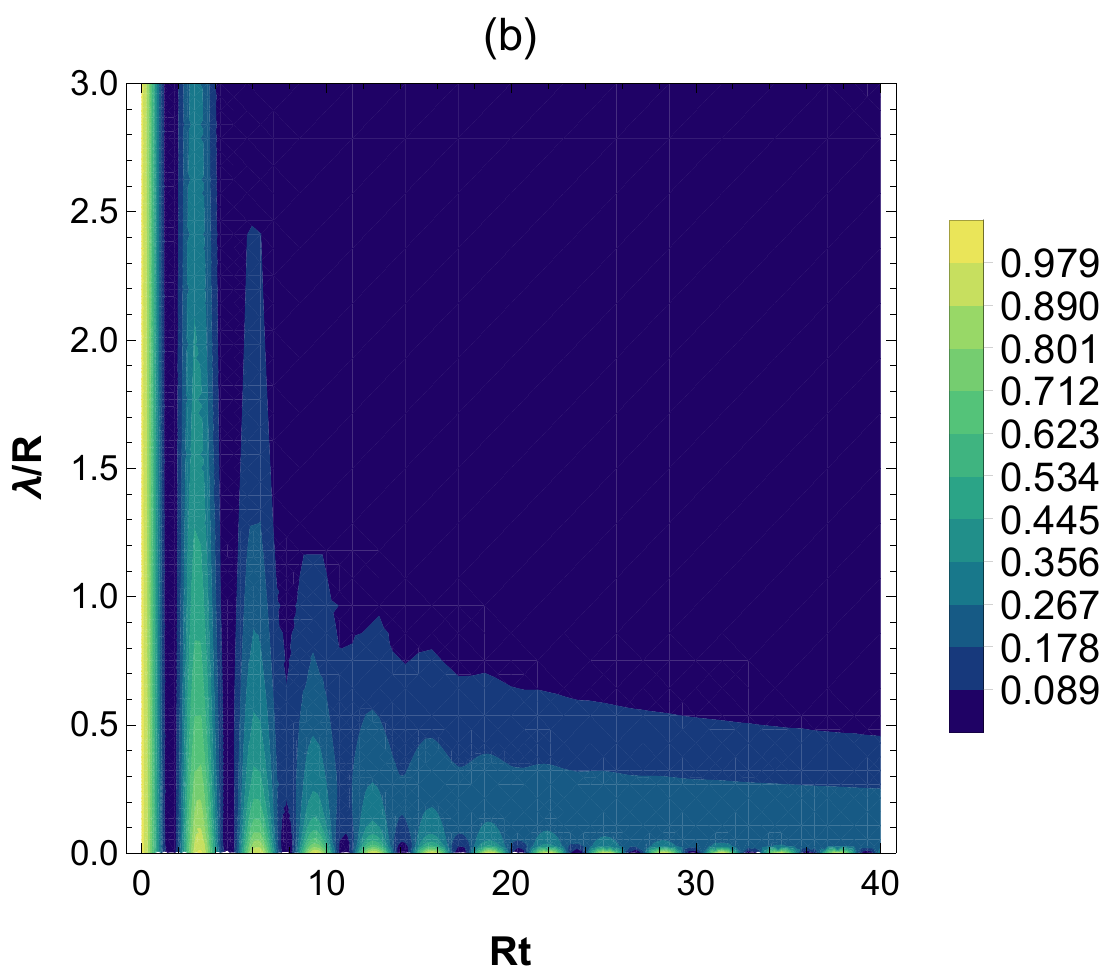}
\parbox{8cm}{\small{\bf Figure 5.}
(Color online)Contour map of quantum Fisher information. (a)$\mathcal{F}_{\phi}$ as a function of $Rt$ and $\frac{\Omega}{R}$ when $\lambda=0.1R$. (b)$\mathcal{F}_{\phi}$ as a function of $Rt$ and $\frac{\lambda}{R}$ when $\Omega=R$.}
\end{center}

In figure 5(a), we plot $\mathcal{F}_{\phi}$ as a function of $Rt$ and $\frac{\Omega}{R}$ when $\lambda=0.1R$. We can find that $\mathcal{F}_{\phi}$ will oscillate damply to a stable value with $\Omega$ increasing, and the larger $\Omega$ corresponds to the faster frequency of oscillation in the strong cavity-reservoir coupling. Figure 5(b) exhibits $\mathcal{F}_{\phi}$ as a function of $Rt$ and $\frac{\lambda}{R}$ when $\Omega=R$. We can know that $\mathcal{F}_{\phi}$ oscillates damply to a stable value only in the strong cavity-reservoir coupling. The smaller the value of $\lambda$ is, the more obvious the oscillation is. Moreover, the evolution behavior of $\mathcal{C}_{l}(t)$ is similar to the $\mathcal{F}_\phi$, we omit it in order to reduce the space.

Therefore, both of the atom-cavity coupling and the cavity-reservoir coupling can effectively protect QFI and QC, the enhancement of parameter estimation precision may occur by adjusting the atom-cavity coupling and the spectral width. Besides, $\mathcal{F}_{\phi}$ is equal to the square of $\mathcal{C}_{l}$ from their stable values, as the same as Eq.~(\ref{EB317}).

\begin{center}
\includegraphics[width=4.2cm,height=3.5cm]{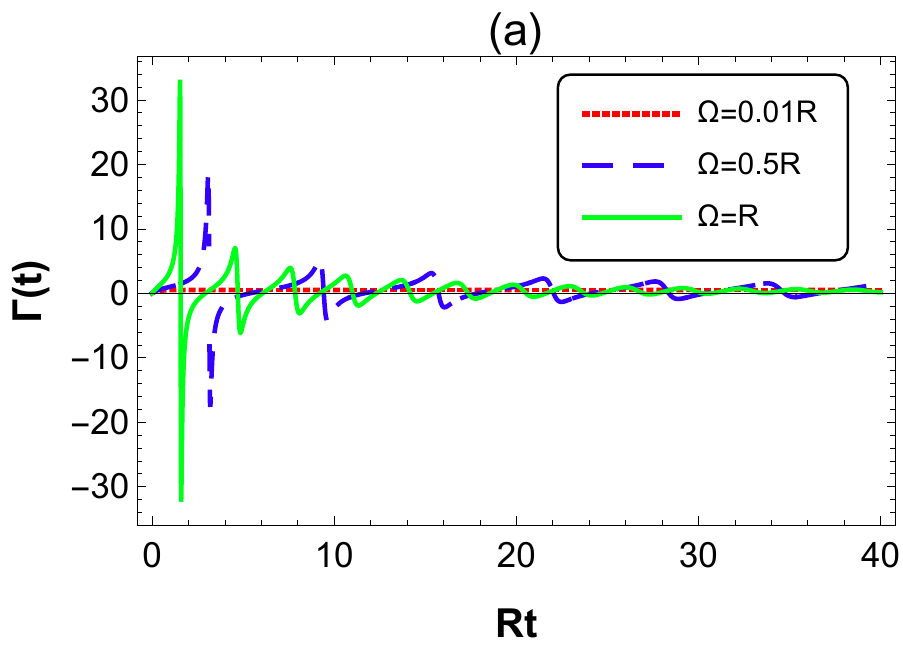}
\includegraphics[width=4.2cm,height=3.5cm]{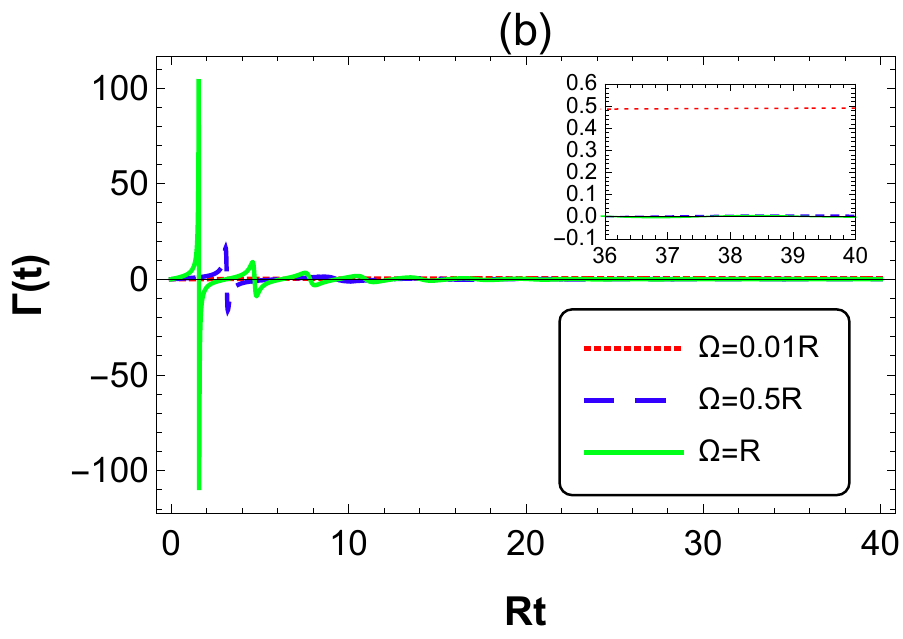}
\parbox{8cm}{\small{\bf Figure6.}
(Color online)The decoherence rate $\Gamma(t)$ versus $Rt$. $\Omega=0.01R$(red,dotted), $\Omega=0.5R$(blue, dashed) and $\Omega=R$(green, solid). (a)$\Gamma_{t}$ as a function of $Rt$ in the weak cavity-reservoir coupling($\lambda=3R$); (b)$\Gamma_{t}$ as a function of $Rt$ in the strong cavity-reservoir coupling($\lambda=0.1R$). Other parameters are $\phi=0$ and $\theta=\frac{\pi}{2}$.}
\end{center}

In order to explain the results in figure 4 and figure 5, we draw the curve of the decoherence rate $\Gamma(t)$ as a function of $Rt$ for the different $\Omega$ value. From Eqs.~(\ref{EB306}), ~(\ref{EB3101}) and ~(\ref{EB313}), we know that the decoherence rate of the atom $\Gamma(t)$ depends on the coupling $\Omega$, the spectral width $\lambda$ and the dissipative rate $R$. The case $\Gamma(t)>0$ indicates that the information flows irreversibly from the atom to the environment, but the case $\Gamma(t)<0$ shows that the information flows back from the environment to the atom. In the weak cavity-reservoir coupling($\lambda=3R$)(figure 6(a)), if $\Omega\leq0.5\omega_{0}$, the information flows irreversibly from the atom to the environment due to the dissipation of reservoir. $\Gamma(t)$ is always positive and $\mathcal{F}_{\phi}$ and $\mathcal{C}_{l}$ reduce quickly to zero with $Rt$. If $\Omega=R$, the information will be fed back to the atom from the environment. $\Gamma(t)$ changes from positive to negative values and then tends to zero so that $\mathcal{F}_{\phi}$ and $\mathcal{C}_{l}$ reduce and then recover to their stable values. In the strong cavity-reservoir coupling($\lambda=0.1R$)(figure 6(b)), if $\Omega\geq0.5R$, the information will flow back from the environment to the atom due to the memory and feedback effect of reservoir. $\Gamma(t)$ becomes larger and alternates between positive and negative values with $\Omega$ increasing so that $\mathcal{F}_{\phi}$ and $\mathcal{C}_{l}$ will oscillate significantly with $Rt$. When $\Omega=R$, $\Gamma(t)$ will tend to zero thus $\mathcal{F}_{\phi}$ and $\mathcal{C}_{l}$ tend to their stable values rather than zero.

\section{Conclusion}
In summary, we investigate QFI and QC of an atom in a dissipative cavity, in which the bosonic reservoir has an Ohmic and a Lorentzian spectral densities at zero temperature, respectively. We acquired the analytical solutions of QFI and QC as well as their relationship. For the Ohmic reservoir, both of the atom-cavity coupling and the cavity-reservoir coupling can effectively protect QFI and QC. Especially, QFI and QC will oscillate significantly with time and tend to their stable values when $\Omega=\omega_{0}$. For the Lorentzian reservoir, we observed that both of the atom-cavity coupling and the cavity-reservoir coupling can also effectively protect QFI and QC. In the weak cavity-reservoir coupling, QFI and QC can tend to their stable values only when the atom-cavity coupling is very strong($\Omega=40R$). In the strong cavity-reservoir coupling, QFI and QC can also tend to their stable values when the atom-cavity coupling is smaller($\Omega=R$). That is, the enhancement of parameter estimation precision will occur by adjusting the atom-cavity coupling whether in the weak or in the strong cavity-reservoir coupling. The quantum coherence can augment QFI and can effectively improve the quantum metrology.

\begin{acknowledgments}
This work was supported by the National Natural Science Foundation of China (Grant No 11374096) and the Doctoral Science Foundation of Hunan Normal University, China.
\end{acknowledgments}

\end{document}